\documentclass[english,keywords,amsmath,amssymb]{revtex4}
\usepackage[T1]{fontenc}
\usepackage[latin1]{inputenc}
\usepackage{braket}
\usepackage{babel}%
\usepackage{graphicx}
\usepackage{color}
\usepackage{bm}
\usepackage{longtable}
\usepackage{amsmath}
\usepackage{amsfonts}
\usepackage{dsfont}
\usepackage{amssymb}
\usepackage{hyperref}
\begin{document}
\title{Revealing universal quantum contextuality through communication games}
\author{A. K. Pan \footnote{akp@nitp.ac.in}}
\affiliation{National Institute of Technology Patna, Ashok Rajhpath, Patna 800005, India}

\begin{abstract}
A theory is universal contextual if its prediction cannot be reproduced by an ontological model satisfying both preparation and measurement noncontextuality assumptions. In this report, we first generalize the logical proofs of quantum preparation and measurement contextuality for qubit system for any odd number of preparations and measurements. Based on the logical proof, we derive  testable universally non-contextual inequalities violated by quantum theory. We then propose a class of two-party communication games and show that the average success probability of winning such games is solely linked to suitable Bell expression whose local bound is greater than universal non-contextual bound. Thus, for a given state, even if quantum theory does not exhibit non-locality, it may still reveal non-classicality by violating the universal non-contextual bound. Further, we consider a different communication game to demonstrate that for a given choices of observables in quantum theory, even if  there is no logical proof of preparation and measurement contextuality exist, the universal quantum contextuality can be revealed through that communication game. Such a game thus test a weaker form of universal non-contextuality with minimal assumption. 
\end{abstract}
\maketitle

Two famous no-go theorems play an important role in quantum foundations as they demonstrate how quantum world differs from the classical ones by ruling out the existence of particular kinds of physical models.  Bell's theorem \cite{bell64} is the first that provides a conflict between the quantum theory and a notion of classicality widely known as local realism. The Kochen and Specker (KS) theorem \cite{bell66,ks} proves an inconsistency between the quantum theory and another notion of classicality known as non-contextual realism. Quantum non-locality found its application in device-independent cryptography\cite{ekert, acin,bar13},  certification of randomness \cite{pir10,acin12,acin16}, self-testing \cite{yao,selftest}, certification of dimension of Hilbert space \cite{dim}. On the other hand, quantum contextuality also provides advantage in communication games \cite{spek09,hameedi,ghorai} and in quantum computation\cite{how,rau}. While the demonstration of Bell's theorem requires two or more space-like separated systems, the KS theorem can also be proven for a single system having dimension of the Hilbert space $d\geq 3$. The original KS proof was demonstrated by using 117 projectors for qutrit system. Later, simpler versions and varients of it using lower number of projectors have been provided \cite{cabello,ker,yu,mermin,cabello08,pan10,kly}. 

The traditional KS proof of contextulity has a limited scope of applicability due to the following reasons.  First, along with the assumption of measurement non-contextuality it additionally requires outcome determinism for sharp measurement in an ontological model. Second, it is not the model of arbitrary operational theory, rather specific to quantum theory. Third, it is not applicable to the generalized measurements, i.e., for POVMs.  The traditional notion of KS non-contextuality was generalized by Spekkens\cite{spek05} for any arbitrary operational theory and extended the formulation to the transformation and preparation non-contextuality. 

The ontological model of an operational theory was coherently formulated by Harrigan and Spekkens\cite{hari}. Let there is a set of preparation procedures $\mathcal{P}$, a set of measurement procedures $\mathcal{M}$ and a set of outcomes $\mathcal{K}_{M}$. Given a preparation procedure $P\in \mathcal{P}$  and a measurement procedures $M\in \mathcal{M}$, an operational theory assigns probability $p(k|P, M)$ of occurrence of a particular outcome $k\in \mathcal{K}_{M}$. For example, in quantum theory, a preparation procedure produces a density matrix $\rho$ and measurement procedure is in general described by a suitable POVM $E_k$. The probability of occurrence of a particular outcome $ k $ is determined by the Born rule, i.e., $p(k|P, M)=Tr[\rho E_{k}]$. In this paper we restrict our discussion in a particular operational theory, i.e., in quantum theory. 

In an ontological model of quantum theory, it is assumed that whenever $\rho$ is prepared by a preparation procedure $P\in \mathcal{P}$ a probability distribution $\mu_{P}(\lambda|\rho)$ in the ontic space is prepared, satisfying $\int _\Lambda \mu_{P}(\lambda|\rho)d\lambda=1$ where $\lambda \in \Lambda$ and $\Lambda$ is the ontic state space. The outcome $k$ is distributed as a response function $\xi_{M}(k|\lambda, E_{k}) $ satisfying $\sum_{k}\xi_{M}(k|\lambda, E_{k})=1$ where a POVM $E_{k}$ is realized through a measurement procedure $M\in\mathcal{M}$. A viable ontological model should reproduce the Born rule, i.e., $\forall \rho $, $\forall E_{k}$ and $\forall k$, $\int _\Lambda \mu_{P}(\lambda|\rho) \xi_{M}(k|\lambda, E_{k}) d\lambda =Tr[\rho E_{k}]$.

An ontological model can be assumed to be non-contextual as follows \cite{spek05}. If two experimental procedures are operationally equivalent, they have equivalent representations in the ontological model. 
An ontological model of quantum theory is assumed to be measurement non-contextual if $\forall P \ \:  p(k|P, M)=p(k|P, M^{\prime})\Rightarrow \forall P  \ \ \xi_{M}(k|\lambda, E_{k})=\xi_{M^{\prime}}(k|\lambda, E_{k})$, where $M$ and $M^{\prime}$ are two measurement procedures realizing the same POVM $E_k$.  KS non-contextuality assumes the aforementioned measurement non-contextuality for the sharp measurements along with the deterministic response functions for projectors. Similarly, an ontological model of quantum theory can be considered to be preparation non-contextual if  $\forall M \ \ :  p(k|P, M)=p(k|P^{\prime}, M)\Rightarrow \forall M \ \ \mu_{P}(\lambda|\rho)=\mu_{P^{\prime}}(\lambda|\rho)$, where $P$ and $P^{\prime}$ are two distinct preparation procedures preparing same density matrix $\rho$. In an ontological model of quantum theory, the preparation non-contextuality implies the outcome determinism for sharp measurements \cite{spek05}. Also, any KS proof can be considered as a proof of preparation contextuality but converse does not hold. In this sense, preparation non-contextuality is a stronger notion than traditional KS non-contextuality \cite{leifer}. Very recently, it is also shown \cite{kunj16} that any ontological model satisfying both the assumptions of preparation and measurement non-contextuality  cannot reproduce all quantum statistics, even if the assumption of outcome determinism for sharp measurement is dropped\cite{kunj16}. Experimental test of such an universal non-contextuality has also been provided which are free from idealized assumptions of noiseless measurements and exact operational equivalences \cite{mazurek}.

In this work, we first generalize the universal contextuality proof demonstrated in \cite{mazurek} for any arbitrary odd number of preparations and measurements in a qubit system.  We then propose  a class of communication games played between two spatially separated parties Alice and Bob and show that the average success probability of winning the game is solely dependent on a Bell expression. The local bound of such a Bell expression gets reduced if universal non-contextuality is assumed. Thus, for a given state, even if quantum theory does not violate local realist bound but it may still reveal non-classicality by violating the universal non-contextuality. We further point out that for a given choices of projectors corresponding to a suitable set of dichotomic observables, if there exists a logical proof of preparation contextuality for mixed state then there will be an inherent interplay between the preparation contextuality for mixed and pure states. In other words, assumption of preparation non-contextuality for mixed state in the logical proof automatically assumes preparation contextuality for pure states.A true test of universal contextuality should be free from such inconsistency.  Further, we demonstrate that for a suitable choices of states and observables when there is no logical proof of preparation and measurement contextuality exist, universal quantum contextuality c an still be revealed through a suitable communication game. 

We first encapsulate the notion of preparation and measurement non-contextuality in an ontological model of quantum theory which was first put forwarded by Spekkens \cite{spek05}.

\section{Logical proofs of  measurement and preparation contextuality}: As already mentioned that the KS proof is restricted to deterministic ontological models and does not work for generalized measurement. In other words, KS proof is valid for commuting contexts and thus the Hilbert space dimension needs to be more than two \cite{bell64,ks}. It is pointed out in \cite{spek05} that the natural question should be whether in a non-deterministic ontological model the probabilities of different outcomes for a given $\lambda$ depend on the compatible context instead of commuting context.  Assuming the indeterministic response functions for POVMs (rigoursly justified in \cite{spek14}) measurement contextuality can be demonstrated even for two-dimensional Hilbert space. 
 
Let $M_1, M_2, M_3 \in \mathcal{M}$ are three measurement procedures in quantum theory realizing three non-degenerate dichotomic observables $A_1, A_2, A_3$ respectively. The projectors corresponding to $A_t$ are $P^{\alpha}_{A_t}=\frac{\mathbb{I}+\alpha A_{t}}{2}$ where $t=1,2,3$ and $\alpha\in \{+,-\}$ satisfying
\begin{eqnarray}
\label{mt}
	\mathbb{I}=P_{A_t}^{+}+P_{A_t}^{-}  
\end{eqnarray}
with $P_{A_t}^{+} P_{A_t}^{-}=0$.

Since $Tr[\mathbb{I} \rho]=1$ for  any arbitrary $\rho$, then in an ontological model $\xi(\alpha|\mathbb{I},\lambda)=1$ for every ontic state $\lambda \in \Lambda$. Thus, the response functions follow the equivalent relations
\begin{align}
\label{mtlambda}
	1= \xi_{M_{t}}(+|P_{A_t}^{+}, \lambda)+ \xi_{M_{t}}(-|P_{A_t}^{-},\lambda) \ \ \ \ \text{with} \ \ \
		\xi_{M_{t}}(+|P_{A_t}^{+},\lambda) \xi_{M_{t}}(-|P_{A_t}^{-},\lambda)=0
\end{align}

Now, consider another measurement procedure $M_{\ast}$, satisfying
\begin{align}
\label{mast}
	\left\{\frac{\mathbb{I}}{2}, \frac{\mathbb{I}}{2}\right\}=\left\{\frac{1}{3} \sum_{t=1}^{3}P_{A_t}^{+}, \ \ \frac{1}{3} \sum_{t=1}^{3}P_{A_t}^{-}\right\}
	\end{align}
	where $\frac{\mathbb{I}}{2}$ is a POVM. Note that, this is only possible if $\sum_{t=1}^{3} A_{t}=0$. A simple choice of qubit observables along trine spin axes satisfies this requirement. The corresponding response function for POVMs follow measurement non-contextuality, so that
	\begin{eqnarray}
\label{mastlambda}
\left\{\frac{1}{2},\frac{1}{2}\right\}=\Big\{ \frac{1}{3}\sum_{t=1}^{3}\xi_{M_{\ast}}(+|P_{A_t}^{+},\lambda),\hskip 0.3cm \frac{1}{3}\sum_{t=1}^{3}\xi_{M_{\ast}}(-|P_{A_t}^{-},\lambda)\Big\} 
\end{eqnarray}

If the model is outcome deterministic for sharp measurement then each $\xi(\alpha|P_{A_t}^{\alpha},\lambda)\in \{0,1\}$ and if measurement non-contextual then $\xi(\alpha|P_{A_t}^{\alpha},\lambda)$ in equation (\ref{mtlambda}) remains same as in equation (\ref{mastlambda}). Importantly, no deterministic assignment can satisfy equation (\ref{mastlambda}) and thus needs to be dropped. However, allowing $\xi(\lambda|P_{A_t}^{\alpha})\in [0,1]$ measurement non-contextuality is satisfied. 

It is important to note that the KS proof does not make any reference to the notion of preparation non-contextuality \cite{spek05} which should be the starting point of any ontological model to ensure that the $\lambda$ distribution for operationally equivalent preparations are the same to make the measurement non-contextuality assumption justified for any $\lambda$. 

Let three preparation procedures $P_1, P_2, P_3\in \mathcal{P}$  produce six pure qubit states $\{\rho_{A_t}^{\alpha}\}$ satisfying following three relations 
\begin{align}
\label{pt}
	\frac{\mathbb{I}}{2}=\frac{1}{2}(\rho_{A_t}^{+}+\rho_{A_t}^{-})
\end{align}
 In an ontological model, using convexity property one can write 
\begin{align}
\label{ptlambda}
	\mu_{P_{t}}(\lambda|\frac{\mathbb{I}}{2})=\frac{1}{2}\left(	\mu_{P_{t}}(\lambda|\rho_{A_t}^{+}) +	\mu_{P_{t}}(\lambda|\rho_{A_t}^{-})\right)
\end{align}
which we call trivial preparation non-contextuallity condition. 

The maximally mixed state $\frac{\mathbb{I}}{2}$ can also be prepared by two more preparation procedures $P_4$ and $P_5$ are of the following form
\begin{align}
\label{pnt}
	\frac{\mathbb{I}}{2}=\frac{1}{3}\sum_{t=1}^{3}\rho_{A_t}^{+}; \ \ \ \ \frac{\mathbb{I}}{2}=\frac{1}{3}\sum_{t=1}^{3}\rho_{A_t}^{-}
		\end{align}
Using the convexity property of the $\lambda$ distributions
\begin{eqnarray}
\label{p45}
	\mu_{P_4}(\lambda|\frac{\mathbb{I}}{2})=\frac{1}{3}\sum_{t=1}^{3}\mu_{P_4}(\lambda|\rho_{A_t}^{+})\\
	\nonumber
		\mu_{P_5}(\lambda|\frac{\mathbb{I}}{2})=\frac{1}{3}\sum_{t=1}^{3}\mu_{P_5}(\lambda|\rho_{A_t}^{-}) 
\end{eqnarray}

which we call non-trivial preparation noncontextuality conditions. Without such non-trivial conditions logical proof of preparation contextuality for mixed state cannot be revealed in this example.

It is indistinguishable in quantum theory by any measurement which of the five preparation procedures is used to prepare the mixed state $\mathbb{I}/2$. Equivalently, in a preparation non-contextual model, it can be assumed that $\lambda$ distributions corresponding to the five preparation procedures are the same, i.e., $\mu_{P_1}(\lambda|\frac{\mathbb{I}}{2})=\mu_{P_2}(\lambda|\frac{\mathbb{I}}{2})=\mu_{P_3}(\lambda|\frac{\mathbb{I}}{2})=\mu_{P_4}(\lambda|\frac{\mathbb{I}}{2})=\mu_{P_5}(\lambda|\frac{\mathbb{I}}{2})\equiv\nu(\lambda|\frac{\mathbb{I}}{2})$. We make it specific here by denoting the above assumption as mixed-state preparation non-contextuality. Importantly, for a  logical proof of preparation non-contextuality for mixed state, one requires to assume preparation non-contextuality of $\lambda$ distribution $\mu(\lambda|\{\rho_{A_t}^{\alpha}\})$ corresponding to the pure states\cite{spek05}. Assume that there exist a $\lambda$ for which $\nu(\lambda|\frac{\mathbb{I}}{2})>0$ and for the same $\lambda$ assume $\mu(\lambda|\{\rho_{A_t}^{+}\})>0$. Then by using $\mu(\lambda|\rho_{A_t}^{+}) \mu(\lambda|\rho_{A_t}^{-})=0$, from equation (\ref{p45}) one finds $\mu_{P_5}(\lambda|\frac{\mathbb{I}}{2})=0$ which is in contradiction with the above assignment. Similar contradiction can be found for any assignment of positive probability distribution for any $\lambda$. However, if preparation contextuality for pure states is assumed, i.e., given a $\lambda$ if $\mu(\lambda|\{\rho_{A_t}^{\alpha}\})$ change their support in $P_{t}$ and  $P_{5}$ or $P_{4}$, the preparation non-contextuality for mixed state will be satisfied  in an ontological model. 

There is a fundamental difference between the logical proofs of measurement and preparation contextuality. Measurement non-contextuality for projector and POVMs may be satisfied in an ontological model if determinism is sacrificed. We note here that in Cabello's \cite{cabello} elegant proof using 18 vector can also be shown measurement non-contextual if all the response functions are taken to be $1/4$. On the other hand, in logical proof of preparation contextuality the assumption of preparation non-contextuality for pure states dictates preparation contextuality for mixed states and vice versa. Hence, such a contradiction is within the ontological model without recourse to any operational theory. Thus, if one wishes to propose a true test of the reproducibility of quantum theory by a universal non-contextual model, the assumptions of outcome determinism and inherent contradiction between preparation non-contextuality for pure and mixed states should be avoided. Such an attempt was made in \cite{kunj16,mazurek} through a testable inequality which is verified experimentally\cite{mazurek}.

\section{Universal non-contextual inequality in (3,3) scenario} We now recapitulate the non-contextual inequality in three-preparation and three- measurement scenario (henceforth, (3,3) scenario) \cite{mazurek}. We start from a simple observation. Let a preparation procedure preparing the maximally mixed state $\mathbb{I}/2$ and a measurement of a qubit observable, so that, $\mathbb{I}=P_{A_{t}}^{+}+ P_{A_{t}}^{-}$.  Since $Tr[\frac{\mathbb{I}}{2}. \mathbb{I}]=1$  one can write $\frac{1}{2}Tr[\left(\rho_{A_{t}}^{+}+\rho_{A_{t}}^{-}\right)$  $\left(P_{A_{t}}^{+}+ P_{A_{t}}^{-}\right)]=1$ irrespective of $P$ and $M$. Here $P_{A_{t}}^{\alpha}$ is equal to $\rho_{A_{t}}^{\alpha}$ but to denote preparation we use $\rho_{A_{t}}^{\alpha}$. In (3,3) scenario, the average correlation can be written as
\begin{align}
\label{cqm3}
	({\Delta_3})_{Q}=\frac{1}{6}\sum_{t=1}^{3}\left(\rho_{A_{t}}^{+} P_{A_{t}}^{+}+\rho_{A_{t}}^{-} P_{A_{t}}^{-}\right)
\end{align}
Due to the perfect correlation in quantum theory each $Tr[\rho_{A_{t}}^{\alpha} P_{A_{t}}^{\alpha}]=1$, one has $(\Delta_3)_{Q}=1$. In an ontological model  
\begin{align}
\label{c31}
	{\Delta_3}=\frac{1}{6}\sum_{t =1}^{3}\sum_{\alpha\in \{+,-\}}\sum_{\lambda\in\Lambda} \xi_{M_{t}}(\alpha|P_{A_t}^{\alpha},\lambda) \mu_{P_t}(\lambda|\rho_{A_t}^{\alpha})
\end{align}
 To get perfect correlation, each $\xi_{M_{t}}(\alpha|P_{A_t}^{\alpha},\lambda)$ should produce deterministic outcome. By assuming mixed state preparation non-contextuality $\nu(\lambda|\mathbb{I}/{2})=\frac{1}{2}\left(\mu(\lambda|\rho_{A_t}^{+})+\mu(\lambda|\rho_{A_t}^{-})\right)$  independent of $t$ and by noting that there is an upper bound on each $\xi_{M_{t}}(\alpha|P_{A_t}^{\alpha},\lambda)$ independent of the outcome $\alpha$, one has $\xi_{M_{t}}(\alpha|P_{A_t}^{\alpha},\lambda)\leq \eta(P_{A_t},\lambda)$. Equation (\ref{c31}) can then be written as

\begin{align}
({\Delta_3})_{unc}\leq \stackrel{max}{\lambda\in\Lambda}\Big(\frac{1}{3}\sum_{t \in \{1,2,3\}}  \eta(P_{A_t},\lambda)\Big)
	\end{align}
If the response functions are constrained by equations (\ref{mtlambda})  and (\ref{mastlambda}), we have $\left( \eta(P_{A_1},\lambda), \eta(P_{A_2},\lambda),\eta(P_{A_3},\lambda)\right)=(1,1/2,1)$. Such an indetrministic assignment in an universal non-contextual model provide  $({\Delta_3})_{unc}\leq 5/6$. Thus, an universal non-contextual model cannot reproduce quantum statistics. This result is verified in a recent experiment \cite{mazurek}. 

We shall shortly generalize the above proof for a qubit system for any arbitrary odd number of observables. Before that,  we provide a scheme to reveal quantum universal contextuality by using a two-party communication game as a tool.
\section{A communication game in (3,3) scenario}

 Let Alice and Bob are two parties having input $x\in \{1,2,3\}$ and $y\in \{1,2,3\}$ respectively and their respective outputs are $a\in\{-1,1\}$ and $b\in \{-1,1\}$. The wining rule is the following;  if $x=y$ the outputs satisfies $a\neq b$ and if $x\neq y $ the outputs satisfies $a=b$. Let Alice and Bob share a maximally entangled state and the input $x\in \{1,2,3\}$ corresponds to the observables ${A_1, A_2, A_3}$ and similarly $y\in \{1,2,3\}$ corresponds to $B_1, B_2, B_3$. Alice measures one of $A_1, A_2, A_3$ to produce six pure qubit states $\{\rho_{A_t}^{\alpha}\}$ satisfying the relation in equations (\ref{pt}) and (\ref{pnt}). The average success probability can be written as 
\begin{eqnarray}
\label{avprob}
	\mathbb{P}_{3,3}=\frac{1}{9}\Big[\sum_{x,y=1}^{3} \left(P(a\neq b| x=y) +P(a = b| x\neq y) \right)\Big]
\end{eqnarray}
Writing $P(a , b|x,y)$ as a moment average $P(a , b|x,y)=\frac{1}{4}[1+a\langle x \rangle+b \langle y\rangle +ab \langle xy \rangle]$, it can shown that $\mathbb{P}_{3,3}$ is solely dependent on a Bell-like expression $\beta_{3,3}$ is given by 
\begin{eqnarray}
\label{avprob}
	\mathbb{P}_{3,3}=\frac{1}{2}\left[1+\frac{\langle \beta_{3,3}\rangle}{9 }\right]
\end{eqnarray}
where 
\begin{eqnarray}
\label{beta33}
\nonumber
	\beta_{3,3}&=& (-A_1 +A_2 +A_3)\otimes B_1 + (A_1 -A_2 +A_3)\otimes B_2 \\
		&+&  (A_1 +A_2 -A_3) \otimes B_3
\end{eqnarray}

A simple choices of Alice's observables $A_1=\sigma_{z}$, $A_2=\frac{\sqrt{3}}{2}\sigma_{x}-\frac{1}{2}\sigma_{z}$ and $A_2=\frac{-\sqrt{3}}{2}\sigma_{x}-\frac{1}{2}\sigma_{z}$  and Bob's observables $B_1=-A_1$, $B_2=-A_{2}$ and $B_{3}=-A_{3}$ provide  the maximum quantum value   $(\beta_{3,3})_{Q}^{max}=6$. This in turn fixes the maximum average success probability in quantum theory is given by
\begin{eqnarray}
	(\mathbb{P}_{3,3})_{Q}\leq\frac{1}{2}\big(1+\frac{2}{3}\big)\approx 0.833
\end{eqnarray}
In two-party, two-outcome Bell scenario, the assumption of preparation non-contextuality is equivalent to locality \cite{barrett}. In such a case, $(\beta_{3,3})_{local}\leq 5$ and $	(\mathbb{P}_{3,3})_{local}=7/9\approx 0.777$. 
This is obtained by assuming that $B_1$, $B_2$ and $B_3$ may take any value in between $-1$ and $+1$. But, given the observable choices in quantum theory that maximimizes 	$(\mathbb{P}_{3,3})_{Q}$ the Bob's observables satisfy the relation given by equation (\ref{mast}) and consequently $B_1+ B_2 + B_3=0$ has to be satisfied. In ontological model equation (\ref{mastlambda}) needs to be satisfied which is measurement-noncontextuality assumption. Thus, in  an universal non-contextual model $(\beta_{3,3})_{unc}\leq 4$ and  the average success probability is given by
\begin{eqnarray}
	(\mathbb{P}_{3,3})_{unc}\leq \frac{1}{2}\big(1+\frac{4}{9}\big)\approx 0.722
\end{eqnarray}
Interestingly, the values of $(\mathbb{P}_{3,3})\in [0.722,0.777]$ do not reveal the non-classicality in the form of quantum nonlocality but in the form of quantum universal contextuality.\\

\section{Generalization for $(n,n)$ scenario for odd $n$} We generalize the universal quantum contextuality proof for any arbitrary preparation and measurements, i.e., the $(n,n)$ scenario. Consider the following $n$ (odd) number of observables  $A_{n,1}=\sigma_{z}$, $ 	\{A_{n, i}\}_{i=2,...\frac{n-1}{2}}=\alpha_{i}\sigma_{x} -\beta_{i}\sigma_{y} - \frac{\sigma_{z}}{(n-1)}$ and $	\{A_{n, j}\}_{j=\frac{n+1}{2}, ... n}=-\alpha_{j}\sigma_{x} +\beta_{j}\sigma_{y} - \frac{\sigma_{z}}{(n-1)}$ with $\alpha_{i}=-\alpha_{j}$, $\beta_{i}=-\beta_{j}$ and $\alpha_{i}^{2} +\beta_{i}^{2} +\frac{1}{n-1}=1$. In quantum theory, such choices of observables satisfy $A_{n,1} +\sum_{i=2}^{\frac{n-1}{2}} A_{n, i}+\sum_{j=\frac{n+1}{2}}^{n} A_{n, j}=0$ and consequently the corresponding projectors satisfy the relation 
\begin{align}
\label{nneq}
	\frac{1}{n}\left(P_{A,1}^{+} +\sum_{i=2}^{\frac{n-1}{2}} P_{A_{n, i}}^{+}+\sum_{j=\frac{n+1}{2}}^{n}  P_{A_{n, j}}^{+}\right)= \frac{\mathbb{I}}{2}
\end{align}
This relation will provide the non-trivial preparation and measurement non-contextual assumptions along with the trivial assumptions originate from $\mathbb{I}=P_{A_{n,t}}^{+} + P_{A_{n,t}}^{-}$ where $t=1,2...n$. A logical proof of preparation and measurement contextuality can also be demonstrated. Now, in $(n,n)$ scenario, the average correlation can be written as
\begin{align}
	\Delta_{n} =\frac{1}{2n}\sum_{t=1}^{n}\sum_{\alpha\in \{+,-\}} p(\alpha|P_{A_{n,t}}^{\alpha}, P_{A_{n,t}}^{\alpha})
\end{align}
In quantum theory, $(\Delta_n)_{Q}=1$. Following the approach adopted earlier, the average correlation in ontological model is 
\begin{eqnarray}
\Delta_{n} &&\leq\frac{1}{n}\sum_{t=1}^{n}\eta(P_{A_{n,t}}, \lambda)\left(\frac{1}{2}\sum_{\alpha\in \{+,-\}} \mu(\lambda|\rho_{A_{n,t}}^{\alpha}) \right)\\
\nonumber
&=& \frac{1}{n}\sum_{t=1}^{n}\eta(\lambda|P_{A_{n,t}})\nu(\lambda|\mathbb{I}/2)
\end{eqnarray}
According to the mixed state preparation non-contextuality assumption $\nu(\lambda|\mathbb{I}/{2})=\frac{1}{2}\Big(\mu(\lambda|\rho_{A_{n,t}}^{+})+\mu(\lambda|\rho_{A_{n,t}}^{-})\Big)$ which is independent of $t$. By noting that it is only relevant if $\lambda$ is in the support of $\nu(\lambda|\frac{\mathbb{I}}{2})$, we have
\begin{eqnarray}
\label{nnn}
	(\Delta_{n})_{unc} \leq \stackrel{max}{\lambda\in\Lambda}\left(\frac{1}{n}\sum_{t=1}^{n}\eta(\lambda|P_{A_{n,t}})\right) 
	\end{eqnarray}
Using equivalent representation of equation (\ref{nneq}) in a measurement non-contextual ontological model and assigning  indeterminstic values of the response function, one has 	$(\Delta_{n})_{unc}\leq 1-\frac{1}{2n}$. Thus, $(\Delta_{n})_{unc} <(\Delta_{n})_{Q}$. However, when $n$ is very large 	$(\Delta_{n})_{unc} \approx (\Delta_{n})_{Q}$. This is due to the fact that  $\{\eta(\lambda|P_{A_{n,t}})\}$ contains only one indeterminstic response function and $(2n-1)$ dsterminstic values out of $2n$ assignments.  
	
Next, we generalize the communication games for $(n,n)$ scenario where $n$ is odd. Let Alice and Bob receive inputs $x\in \{1,2,...n\}$ and $y\in \{1,2,....n\}$ respectively and their respective outputs are $a\in\{-1,1\}$ and $b\in \{-1,1\}$. The winning rule remains same, i.e.,   if $x=y$ then $a\neq b$ and  if if $x\neq y$ then $a=b$ . The average success probability is given by 
\begin{eqnarray}
\nonumber
	\mathbb{P}_{n,n}=\frac{1}{n^2}\Big[\sum_{x,y =1}^{n}\left(P(a\neq b|x,y; x=y)+ P(a = b|x,y; x\neq y) \right)\Big]
\end{eqnarray}
 which can be cast as   
\begin{eqnarray}
	\mathbb{P}_{n,n}= \frac{1}{2}+\frac{\beta_{n}}{2n^2}
\end{eqnarray}
where $\beta_{n}$ is the Bell expression is given by

\begin{align}
	\beta_{n,n}=\sum_{x,y=1 ; x\neq y}^{n} \langle A_x B_y\rangle -\sum_{x,y=1 ; x= y}^{n} \langle A_x B_y\rangle
\end{align}
Given the choices of observables satisfy equation (\ref{nneq}), in quantum theory $(\beta_{n,n})_{Q}=2n$ providing quantum success probability
\begin{align}
	(\mathbb{P}_{n,n})_{Q}=\frac{1}{2}+\frac{1}{n}
\end{align}
For a universal non-contextual model satisfying the equivalent representation of the equation \ref{nneq}) in the ontological model, we have $(\beta_{n,n})_{unc}=2n-2$ which provides the average success probability in an universal non-contextual model 
\begin{align}
	(\mathbb{P}_{n,n})_{unc}=\frac{1}{2}+\frac{1}{n}-\frac{1}{n^2}
\end{align}
 Since $(\mathbb{P}_{n,n})_{Q}>(\mathbb{P}_{n,n})_{unc}$ for any arbitrary $n$,  the universal quantum contextuality is revealed through the communication game. 

We make a few comments on the above proofs of universal quantum contextuality. For the above special choices of observables, whenever preparation non-contextuality for mixed state in an ontological model is assumed, one may argue that  the preparation contextuality for pure state is automatically installed within the onlogical model. One may then say that inequality (\ref{nnn}) for any odd $n$ does not provide a true test of universal quantum contextuality. This is due to the fact that assumption of preparation non-contextuality leads us to assume preparation contextuality for pure states and vice versa. However, this feature was not required to be used in the derivation of the inequality (\ref{nnn}). In the Appendix, we provide an example of $(4,4)$ scenario where no logical proof of preparation and measurement contextuality can be demonstrated and no contradiction with quantum theory through the inequality similar to equation (\ref{nnn}) can be shown. It would then be interesting if the violation of universal non-contextuality can be shown when there is no logical proof possible. We provide such a proof through the communication game in $(3,4)$ scenario.

\section{Communication games in (4,3) and (3,4) scenarios} 
Before presenting the communication game in $(3,4)$ scenario we first demonstrate the communication game in $(4,3)$ scenario. The  game in (4,3) scenario has close resemblance with the $3$ to $1$  parity-oblivious random access code. In \cite{spek09,ghorai}, it was shown that how quantum preparation contextuality powers $3$ to $1$  random access code.    

In (4,3) scenario, Alice and bob measures four and three dichotomic observables respectively. The winning rule is the following; if $x+y=5$ output requires $a\neq b$ and if $x+y\neq 5$ output requires $a= b$. The average success probability can be written as 
\begin{eqnarray}
\label{avprob}
	\mathbb{P}_{4,3}=\frac{1}{12}\Big[\sum_{x=1}^{4} \sum_{y=1}^{3}\left(P(a\neq b|x,y; x+y=5) + P(a = b|x,y; x+y\neq 5) \right)\Big]
\end{eqnarray}
which can be recast as 
\begin{align}
\label{pelegant}
	\mathbb{P}_{4,3}=\dfrac{1}{2} + \dfrac{\langle{\beta_{4,3}}\rangle}{24} 
\end{align}
where $ \beta_{4,3} = (A_1 + A_2 + A_3 - A_4)\otimes B_1 + (A_1 + A_2 - A_3 + A_4)\otimes B_2  + (A_1 - A_2 + A_3 + A_4)\otimes B_3$ is known as Gisin's elegant Bell expression.  The maximization of $ \beta_{4,3}$  in turn provides the optimal success probability of the (4,3) game. In quantum theory, if the  projectors with $+1$ eigenvalues corresponding to Alice's observables forms a four-outcome SIC-POVM and Bob's observables are mutually unbiased basis, then $( \beta_{4,3})^{opt}_{Q} = 4\sqrt{3}$ \cite{gisin}. Then, $(\mathbb{P}_{4,3})_{Q}^{opt}=(1 +1/\sqrt{3})/2\approx 0.788$. A model assuming only trivial non-contextuality is in fact a local model. In such a case, $(\beta_{4,3})_{local} \leq 6$ and $	(P_{4,3})_{local}= (1+1/2)/2= 0.75$.   However, if one assumes additional non-trivial non-contextuality assumption (given in detail in Appendix), a constraint $ A_1 = A_2+A_3+A_4 $ on Alice observables needs to be satisfied. The operational equivalence in quantum theory dictates to assume such non-contextuality, i.e., functional relation between $A_{t}$ in the ontological model. Under such constraint of non-contextuality, the preparation-noncontextual bound on Bell expression  $(\beta_{4,3})_{pnc}\leq 4$ and the average success probability is given by $	(\mathbb{P}_{4,3})_{pnc}= (1+1/3)/2\approx 0.666$.  The importance of $(4,3)$ game example is that the addition of non-trivial non-contextuality assumption along with the trivial ones do not provide the logical contradiction of preparation non-contextuality for pure or mixed state. This thus truly the test of preparation contextuality free from logical inconsistency. 

Due to the symmetry, the communication game in $(4,3)$ scenario can easily be converted to $(3,4)$ one by swapping the role of Alice and Bob's observables. In that case it becomes a proof of universal contextuality. Let Alice performs measurement of three mutually unbiased bases. So that, there is no functional relation between Alice three observables and thus no non-trivial preparation non-contextuality assumption can be made. In other words, no logical proof of preparation contextuality can be shown for such choices of observables. For Bob's choices of observables $\{B_t\}$ (where $t=1,2,3,4$) along with the trivial conditions $\frac{\mathbb{I}}{2}=\frac{1}{2}\Big(P_{B_t}^{+}+ P_{P_t}^{-}\Big)$ the POVM measurements are (the non-trivial conditions)  
\begin{eqnarray}
\label{mnc4nt}
\frac{\mathbb{I}}{2}=\frac{1}{4}\left(P_{B_1}^{+}+\sum_{t=2}^{4} P_{B_t}^{-}\right); \ \ \ \ \ \  \frac{\mathbb{I}}{2}= \frac{1}{4}\left(P_{B_1}^{-}+\sum_{t=2}^{4}P_{B_t}^{+}\right)
\end{eqnarray} 
Here, each $P_{A_t}^{\alpha}=(1+\alpha B_{t})/2$ is a rank one projector corresponding to  dichotomic observables $\{B_{t}\}$ having eigenvalues $\alpha\in \{1,-1\}$. 
To satisfy the relations in equation (\ref{mnc4nt}) one requires the functional relation $B_1=B_2 +B_3 +B_4$ to be satisfied. This is valid even for deterministic and measurement non-contextual values of the response functions. Hence, no logical proof of measurement non-contextuality can be demonstrated irrespective of the nature of the response function. 
 
By keeping the winning rule same, in $(3,4)$ scenario, the success probability is given by
\begin{eqnarray}
\label{avprob}
	\mathbb{P}_{3,4}=\frac{1}{12}\Big[\sum_{x =1}^{3}\sum_{y =1}^{4} 	\Big[P(a\neq b|x,y; x+y=5) +P(a = b|x,y; x+y\neq 5) \Big]\Big]
\end{eqnarray}
which can be re-written in the following form 
  \begin{align}
\label{pelegant}
	\mathbb{P}_{3,4}=\dfrac{1}{2} + \dfrac{{\langle {\beta_{3,4}}}\rangle}{24} 
\end{align}
where $\beta_{3,4} = \beta_{4,3}$ and maximum quantum values is $4\sqrt{3}$ for the choices of the observables given in the Appendix. Thus the universal non-contextual bound for $(\beta_{3,4})_{unc}\leq 4$ providing the success probability  $	(\mathbb{P}_{3,4})_{unc}\leq 0.666$ and quantum theory violates this bound. Hence, we have provided a proof to reveal universal quantum contextuality where there is no logical proof of preparation or measurement contextuality exist for the set of choices of the observables in quantum theory. This thus can be considered as a true test of universal quantum contextuality.
\section{Summary and discussions}
We discussed the notions of preparation and measurement non-contextuality in an ontological model of quantum theory and generalized it for any arbitrary odd number of preparation and measurement scenario.  We then derived the universal non-contextual inequality for any odd number of preparation and measurement scenario and demonstrated the quatum violation of them. Further, schemes are proposed to test universal quantum contextuality by using two-party communication games as tool. It is shown that the average success probability of winning such a game is solely dependent on a suitable Bell expression. For two-party and dichotomic measurements the trivial preparation non-contextuality is equivalent to locality assumption. Interestingly, the local bound of such a Bell expression gets reduced if non-trivial preparation and measurement non-contextuality conditions i.e., universal non-contextuality is assumed. Thus, for a given state, even if quantum theory does not violate local realist bound, there is a possibility to reveal non-classicality through the violation of universal non-contextuality. 

We have also pointed out the subtleties involved in the logical proof of quantum preparation and measurement contextuality for a suitable set of pure qubit states corresponding to a suitable set of dichotomic observables. If there exists a logical proof of preparation contextuality for mixed state then there will be an inherent interplay between the preparation contextuality for mixed and pure states. In other words, in order to impose the assumption of preparation non-contextuality for a mixed state in an ontological model,  the assumption of preparation contextuality for pure states constituting the relevant mixed state requires to be assumed. Such contradiction appears within the framework of concerned ontological model devoid of any operational throy. If one wishes to truly test the universal non-contextuality such inconsistency needs to be avoided. We demonstrated that for a suitable choices of states and observables when is no logical proof of preparation and measurement contextuality can exist, universal quantum contextuality can still be revealed through a suitable communication game. All the proofs provided in this paper is derived for qubit system and  experimentally testable using existing technology. However, the precise operational equivalence has to be ensured in the real experiments  following a recently developed approach \cite{pusey,mazurek} . 

\section*{Acknowledgements}
 AKP acknowledges the support from the project DST/ICPS/QuST/Theme-12019/4.

\begin{widetext}
\appendix
\section{The $(4,4)$ scenario}

 We show that in four-preparation and four-measurement scenario (the $(4,4)$ scenario) no logical contradiction of preparation and measurement non-contextuality can be demonstrated. We further show that in such case no contradiction with quantum theory through the inequality similar to equation (\ref{nnn}) can be demonstrated.  We then propose a communication game similar to $(n,n)$ scenario to show that there is no contradiction as well.

Let four preparation procedures $\{P_t\}\in \mathcal{P}$ where $t=1,2,3,4$  realize the observables ${A_{t}}$ and produce eight pure qubit states $\{\rho_{A_t}^{\alpha}\}$ satisfying following four relations 
\begin{align}
\label{pt4}
	\frac{\mathbb{I}}{2}=\frac{1}{2}(\rho_{A_t}^{+}+\rho_{A_t}^{-})
\end{align}
 In an ontological model, the $\lambda$ distribution can be written as  
\begin{align}
\label{ptlambda}
	\mu_{P_{t}}(\lambda|\frac{\mathbb{I}}{2})=\frac{1}{2}\left(	\mu_{P_{t}}(\lambda|\rho_{A_t}^{+}) +	\mu_{P_{t}}(\lambda|\rho_{A_t}^{-})\right)
\end{align}
which are the trivial preparation non-contextuallity condition. 

If eight qubit projectors $\{\rho_{A_t}^{\alpha}\}$ are obtained from Alice's measurement of four observables are $A_1 = (\sigma_x + \sigma_y + \sigma_z )/\sqrt{3}$, $A_2 = (\sigma_x + \sigma_y - \sigma_z )/\sqrt{3}$, $A_3 = (\sigma_x - \sigma_y + \sigma_z) /\sqrt{3}$ and $A_4 = (-\sigma_x + \sigma_y + \sigma_z )/\sqrt{3}$. Then the maximally mixed state $\frac{\mathbb{I}}{2}$ can also be prepared by two more preparation procedures $P_5$ and $P_6$ are of the following form
\begin{align}
\label{pnt4}
	\frac{\mathbb{I}}{2}=\frac{1}{4}\Big(\rho_{A_1}^{+}+\sum_{t=2}^{3}\rho_{A_t}^{-}\Big); \ \ \ \ \frac{\mathbb{I}}{2}=\frac{1}{4}\Big(\rho_{A_1}^{-}+\sum_{t=2}^{3}\rho_{A_t}^{+}\Big)
		\end{align}
 If ontological model is preparation non-contextual for mixed states, then $\mu_{P_{1}}(\lambda|\frac{\mathbb{I}}{2})=\mu_{P_{2}}(\lambda|\frac{\mathbb{I}}{2})=\mu_{P_{3}}(\lambda|\frac{\mathbb{I}}{2})=\mu_{P_{4}}(\lambda|\frac{\mathbb{I}}{2})=\mu_{P_{5}}(\lambda|\frac{\mathbb{I}}{2})=\mu_{P_{6}}(\lambda|\frac{\mathbb{I}}{2})=\nu(\lambda|\frac{\mathbb{I}}{2})$. One can then write  
\begin{subequations}
\begin{eqnarray}
\label{pnc4logical1}
\nu(\lambda)&=&\frac{1}{2}\Big(\mu_{P_{t}}(\lambda|\rho_{A_t}^{+})+ \mu_{P_{t}}(\lambda|\rho_{A_t}^{-})\Big)\\
\label{pnc4logical2}
&=&\frac{1}{4}\left(\mu_{P_{5}}\left(\lambda|\rho_{A_1}^{+}\right)+\sum_{t=2}^{4}\mu_{P_{5}}\left(\lambda|\rho_{A_t}^{-}\right)\right)\\
\label{pnc4logical3}
&=&\frac{1}{4}\left(\mu_{P_{5}}\left(\lambda|\rho_{A_1}^{-}\right)+\sum_{t=2}^{4}\mu_{P_{5}}\left(\lambda|\rho_{A_t}^{+}\right)\right)
\end{eqnarray}
\end{subequations}
Let us now consider the values of those epistemic states for a fixed $\lambda$. Since $\rho_{A_{t}^{+}}$ is orthogonal to $\rho_{A_{t}^{-}}$, then there is no common $\lambda$ in the support of both $\mu(\lambda|\rho_{A_{t}^{+}})$ and $\mu(\lambda|\rho_{A_{t}^{+}})$, so that $\mu(\lambda|\rho_{A_{t}^{+}}) \mu(\lambda|\rho_{A_{t}^{-}})=0$.

If for a given $\lambda$, we have $\mu_{A_{t}}(\lambda|\rho_{A_{1}^{+}})$, $\mu(\lambda|\rho_{A_{2}^{-}})$, $\mu(\lambda|\rho_{A_{3}^{-}})$ and $\mu(\lambda|\rho_{A_{4}^{-}})$ are zero, then from equation(\ref{pnc4logical2}) we have $\nu(\lambda)$ is zero which contradicts with all other conditions. But, if one assumes the $\lambda$ is in the support of $\mu(\lambda|\rho_{A_{1}^{+}})$, $\mu(\lambda|\rho_{A_{2}^{-}})$, $\mu(\lambda|\rho_{A_{3}^{+}})$ and $\mu(\lambda|\rho_{A_{4}^{+}})$, then equations (\ref{pnc4logical1}- \ref{pnc4logical3})) can be written as
\begin{subequations}
\begin{eqnarray}
\label{pnc4l1}
\nu(\lambda)&=&\frac{1}{2}\Big(\mu_{P_{3}}(\lambda|\rho_{A_3}^{+})\Big)\\
\label{pnc4l2}
&=&\frac{1}{2}\Big(\mu_{P_{4}}(\lambda|\rho_{A_4}^{+})\Big)\\
\label{pnc4l3}
&=&\frac{1}{4}\Big(\mu_{P_{6}}(\lambda|\rho_{A_3}^{+})+ \mu_{P_{6}}(\lambda|\rho_{A_4}^{+})\Big)
\end{eqnarray} 
\end{subequations}
It can be easily checked from equations(\ref{pnc4l1}-\ref{pnc4l3})  that they are consistent and no logical proof of preparation contextuality for the above scenario.

From equations (\ref{pt4}) and (\ref{pnt4}) in an ontological model, one can write
\begin{subequations}
\begin{eqnarray}
	1&=& \xi_{M_t}(+|P_{A_{t}}^{+})+ \xi_{M_t}(+|P_{A_{t}}^{-})\\
	\frac{1}{2}&=&\frac{1}{4}\left(\xi_{M_5}(+|P_{B_{t}}^{+})+\sum_{t=2}^{4}\xi_{M_5}(+|P_{t}^{-})\right); \ \ \ \ \frac{1}{4}\left(\xi_{M_6}(+|P_{B_{t}}^{-})+\sum_{t=2}^{4}\xi_{M_6}(+|P_{t}^{+})\right)
\end{eqnarray}
\end{subequations}
It can be easily shown that all the response functions may take determinstic value and non-contextual. 

Next, the average correlation for $(4,4)$ scenario in quantum theory can be written as
\begin{align}
	\Delta_{4,4}=\frac{1}{8}\sum_{t= 1}^{4}\sum_{\alpha\in \{+,-\}} p(\alpha|\rho^{\alpha}_{t}, P_{A_t}^{\alpha})
\end{align}
Note that in quantum theory, $(\Delta_{4,4})_{Q}=1$. In an ontological model as
\begin{align}
\label{c44}
	\Delta_{4,4}=\frac{1}{8}\sum_{t= 1}^{4}\sum_{\alpha\in \{+,-\}}\sum_{\lambda\in\Lambda} \xi(\alpha|P^{\alpha}_{A_t},\lambda) \mu(\lambda|\rho^{\alpha}_{A_t})
\end{align}

 In order to get perfect average correlation, the each term in the equation(\ref{c44}) needs to be perfectly correlated, implying that every response function $\xi(\alpha|P^{\alpha}_{A_t},\lambda)$ should produce deterministic outcome. Now, applying trivial preparation non-contextuality as in \cite{mazurek}, we can write 

\begin{align}
(\Delta_{4,4})_{unc}\leq \stackrel{max}{\lambda\in\Lambda}\Big(\frac{1}{4}\sum_{t=1}^{4}  \eta(P_{A_t},\lambda)\Big)
\end{align}
The quantity $\eta(P_{A_t},\lambda)$ can be maximized with the condition given in equations (\ref{pt4}) and (\ref{pnt}). It is simple to show that  $\Big( \eta(M_{1},\lambda), \eta(M_{2},\lambda),\eta(M_{3},\lambda),\eta(M_{4},\lambda)\Big)=\Big(1,1,1,1\Big)$. We then have , $(\Delta_{4,4})_{unc}=(\Delta_{4,4})_{Q}=1$. Thus, in this case universal non-contextual model can reproduce the perfect predictability of quantum theory.

\section{A communication game in $(4,4)$ scenario:}
Let Alice and Bob are two parties having input $x\in \{1,2,3,4\}$ and $y\in \{1,2,3,4\}$ respectively and outputs are $a\in\{-1,1\}$ and $b\in \{-1,1\}$. The wining rule is the following. If $x=y$ the outputs satisfies $a\neq b$ and if $x\neq y$ the outputs satisfies $a=b$. Let the input $x\in \{0,1,2,3\}$ corresponds to ${A_1, A_2, A_3, A_{4}}$ and $y\in \{0,1,2\}$ corresponds to $B_1, B_2, B_3$ with $A_i=B_i$. The average success probability can be written as

\begin{eqnarray}
\label{avprob}
	\mathbb{P}=\frac{1}{16}\Big[\sum_{x,y =1}^{4}\Big[P(a\neq b|x,y; x=y) +P(a = b|x,y; x\neq y) \Big]\Big]
\end{eqnarray}

which can then be cast as

\begin{align}
\label{pelegant}
	\mathbb{P}_{4,4}=\frac{1}{2} + \frac{\langle \beta_{4}\rangle}{32} 
\end{align}
where 
\begin{align}
\label{elegant}
\beta_{4,4} = &A_1\otimes (-B_1 + B_2 + B_3 + B_4) + A_2\otimes(B_1 - B_2 + B_3 + B_4)+ A_3\otimes (B_1 + B_2 - B_3 + B_4)+ A_4\otimes (B_1 + B_2 + B_3 - B_4)
\nonumber
\end{align} 
It can be shown that $(\beta_{4,4})_{unc}=(\beta_{4,4})_{Q}\leq 8$. Universal quantum contextuality cannot be demonstrated throgh the $(4,4)$ game considered here.
\end{widetext}

\end{document}